# $Cs_3V_9Te_{13}$: A New Vanadium-Based Material with a Reuleaux-Triangle-Like Lattice and a Possible Phase Transition near 48 K

*Zhen Zhao[1#], Jianping Sun[1,2#], Xin-Wei Yi[3#], Ruwen Wang[1,2#], Lin Zhu[1], Tong Liu[2], Haisen Liu[1], Hui Guo[1,2], Wu Zhou[2], Jinguang Cheng[1,2,\*], Gang Su[3\*], Haitao Yang[1,2\*], and Hong-Jun Gao[1,2\*]*

[1] Beijing National Center for Condensed Matter Physics and Institute of Physics, Chinese Academy of Sciences, Beijing 100190, PR China

[2] School of Physical Sciences, University of Chinese Academy of Sciences, Beijing 100190, PR China

[3] Institute of Theoretical Physics, Chinese Academy of Sciences, Beijing 100190, China

[#]These authors contributed equally to this work

[*]Correspondence to: hjgao@iphy.ac.cn, htyang@iphy.ac.cn, sugang@itp.ac.cn, jgcheng@iphy.ac.cn

**Exploring and synthesizing materials with new crystal structures provides an important route to discovering exotic quantum phenomena. However, materials with unconventional lattice geometries remain largely unexplored. Here, we report the discovery of a new vanadium-based material, $Cs_3V_9Te_{13}$, featuring a Reuleaux-triangle-like lattice. Electrical transport and magnetic measurements consistently reveal an anomaly near 48 K, and this feature shows little sensitivity to the applied magnetic field. A corresponding anomaly is also observed in the Hall coefficient near 48 K, indicating a marked change in the carrier response. In addition, temperature-dependent x-ray diffraction results indicate no obvious structural change across 48 K. Taken together, these results suggest that the anomaly is not induced by the structural transition, but associated to a possible electronic and/or magnetic phase transition. High-pressure transport measurements and first-principles calculations further reveal a highly tunable electronic state in $Cs_3V_9Te_{13}$, with the kagome-like electronic feature and pressure-suppressed antiferromagnetism. These results demonstrate this material, with its structurally novel Reuleaux-triangle-like lattice, as a new platform for exploring the interplay between nontrivial lattice geometry and emergent physical phenomena.**

The geometry of a crystal lattice plays a central role in shaping electronic structures and emergent quantum states[1]. In condensed-matter systems, geometrically nontrivial lattices often host unusual band dispersions and enhanced many-body instabilities, thereby providing fertile platforms for exploring exotic physical phenomena[2,3]. For example, the kagome lattice has attracted particular attention because it naturally hosts Dirac points, flat bands, and van Hove singularities[4]. Owing to these remarkable electronic characteristics, kagome materials have become an important platform for exploring the magnetism, superconductivity, and intertwined electronic states[5–9].

The vanadium-based kagome metal family $A$V$_3$Sb$_5$ ($A$ = K, Rb, Cs) has attracted widespread attention as a representative kagome system, featuring a distinctive kagome-lattice-derived electronic structure and a variety of novel emergent phenomena[10-15]. Beyond superconductivity and charge-density-wave order, a variety of unconventional behaviors have been reported in CsV$_3$Sb$_5$, including the electronic nematicity[16], switchable chiral transport[17], pair-density-wave order[18], and long-range coherent charge transport in the normal state[19]. The diversity of these observations highlights the strong sensitivity of kagome-derived quantum states to the lattice geometry and chemical environment, and motivates the search for new compounds beyond the AV$_3$Sb$_5$ framework. This tunability has been further demonstrated by newly discovered members of the broader "135" family[20]. For example, CsTi$_3$Bi$_5$ exhibits superconductivity without an accompanying charge-density-wave transition[21,22], while CsCr$_3$Sb$_5$ shows much stronger magnetic and correlation effects[23,24], illustrating that the suitable chemical substitution within the same structural family can profoundly reconstruct both the lattice framework and the resulting electronic ground states.Motivated by this structural and electronic tunability, it is highly desirable to search for new kagome-related compounds beyond the currently known members. In particular, chemical substitution can induce stronger lattice reconstruction, driving kagome-derived frameworks into structurally distinct regimes with new geometric motifs and emergent physical properties.

Here, we report the synthesis of a new layered compound, Cs$_3$V$_9$Te$_{13}$, whose structure can be regarded as a distorted derivative of the CsV$_3$Sb$_5$-type motif. Distinct from the perfect kagome framework, Cs$_3$V$_9$Te$_{13}$ exhibits a highly intricate crystal structure in which the V sublattice forms a Reuleaux-triangle-like network, giving rise to an unusual geometry beyond the conventional "135" family. Such a pronounced structural distortion, together with the modified stoichiometry, is likely associated with the replacement of Sb by Te, which changes both the chemical valence balance and ionic-size matching in the lattice. Moreover, transport, Hall, and magnetic measurements reveal a possible phase transition near 48 K. The anomaly is insensitive to the applied magnetic field and is not accompanied by detectable structural changes in the temperature-dependent x-ray diffraction, suggesting a nonstructural transition of likely electronic or magnetic origin. Under high pressure, the transport behavior exhibits a non-monotonic evolution, indicating that the electronic states of $Cs_3V_9Te_{13}$ are highly tunable under the lattice compression. These results identify $Cs_3V_9Te_{13}$ as a new platform for exploring the interplay between nontrivial lattice geometry and emergent phenomena.

The Reuleaux triangle (Fig. 1a), formed by three equal-radius circular arcs, is a geometrical figure that preserves the threefold rotational symmetry while maintaining a constant width in all directions, and is therefore of fundamental interest in the mathematics and engineering design[25, 26]. In condensed-matter physics, however, lattices with Reuleaux-triangle-like geometry have remained largely unexplored. In this spirit, the lattice shown in Fig. 1b, formed by the interconnected triangular, quadrilateral, and pentagonal units, exhibits the characteristic geometry of a Reuleaux triangle and can therefore be described as a Reuleaux-triangle-like lattice. To gain insight into its electronic properties, we calculate the band structure of this lattice using a nearest-neighbor tight-binding model (Fig. 1c). The hopping parameters are set to $t_1$ = -1 eV, $t_2$ = -0.9 eV, $t_3$ = -0.8 eV, and $t_4$ = -0.7 eV, with their magnitudes assigned based on the sequence of atomic bond lengths. The on-site energies $\varepsilon_1$ and $\varepsilon_2$ are chosen as 1.5 and 1 eV, respectively. Notably, the electronic band structure of this lattice closely resembles that of the kagome lattice, hosting a rich variety of exotic electronic states, including massless Dirac points, van Hove singularities with a high density of states, and flat bands originating from the destructive quantum interference.

$Cs_3V_9Te_{13}$ crystallizes in a layered hexagonal structure with space group P-62m (Fig. 1d). Along the *c* axis, the crystal can be described by a Cs–Te2–VTe1–Te2–Cs stacking sequence, reminiscent of the layered architecture of $CsV_3Sb_5$[11, 13]. The crystal structure consists of a highly complex framework built from several intertwined atomic sublattices. The V atoms form a distinctive two-dimensional network of a Reuleaux-triangle-like lattice (Fig. 1e). Furthermore, the V–Te1 layer is sandwiched between two Te2 layers, in which the Te2 atoms form a distinct sublattice composed of edge-sharing pentagons and triangles (Fig. 1f). Together, these intertwined sublattices give rise to the intricate layered framework, highlighting the structural complexity of $Cs_3V_9Te_{13}$ beyond the conventional kagome motif. Notably, $Cs_3V_9Te_{13}$ is stable under ambient conditions and can be mechanically exfoliated, highlighting its potential as a layered material.

The representative XRD pattern of a $Cs_3V_9Te_{13}$ single crystal only show the (00l) peaks (Fig. 2a), indicating the preferred 00l orientation, consistent with its layered structure. The calculated c-axis lattice parameter is 8.2181 Å, smaller than that of $CsV_3Sb_5$, possibly due to the smaller ionic radius of Te compared with Sb. The inset is an optical image of a representative single crystal with a regular hexagonal morphology. Energy-dispersive spectroscopy (EDS) analysis of bulk $Cs_3V_9Te_{13}$ gives a semi-quantitative atomic ratio of Cs: V: Te ≈ 3.1:9.0:13.0, in good agreement with its nominal stoichiometry. To further verify this novel structure, atomic-resolution high-angle annular dark-field (HAADF) scanning transmission electron microscopy (STEM) measurements were carried out. The Z-contrast image of $Cs_3V_9Te_{13}$ viewed along the [120] projection is shown in Fig. 2c, with the structural model overlaid. The atomic-resolution images clearly reveal the perfect crystalline structure of the $Cs_3V_9Te_{13}$ sample without noticeable structural defects or impurity phases. Furthermore, the experimental STEM bright-field (BF) image is in good agreement with the simulated BF image, further confirming the crystal structure as determined by the atomic-resolution characterization. These results collectively confirm the high crystallinity, phase purity, and well-defined layered structure of $Cs_3V_9Te_{13}$, providing a solid basis for the subsequent investigation of its physical properties.

The physical properties of $Cs_3V_9Te_{13}$ were further examined through the transport and magnetic measurements. As shown in Fig. 3a, the resistance of the $Cs_3V_9Te_{13}$ single crystal exhibits a non-monotonic temperature dependence, with a hump-like feature. As the temperature increases, the resistance increases up to about 100 K and then decreases monotonically. In addition, a weak kink is visible near 48 K, as highlighted in the inset of Fig. 3a. This anomaly is reminiscent of the transport signature associated with the charge-density-wave transition in $CsV_3Sb_5$[11]. To visualize the kink clearly, the temperature derivative of the resistivity, $d\rho/dT$, is plotted in Fig. 3b. There is a pronounced peak at 48 K, corresponding to the kink in the resistance curve. To investigate the evolution of the anomaly under magnetic fields, the field-dependent resistance measurements were carried out. As shown in Fig. 3b, the peak remains clearly visible up to 5 T, with no obvious shift in temperature, indicating that the anomaly near 48 K is robust against the applied magnetic fields.

Hall measurements were performed on $Cs_3V_9Te_{13}$ to investigate its charge transport properties as shown in Fig. 3c. The Hall resistivity shows an almost linear dependence on the magnetic field over the entire measurement range for the temperature. Notably, the slope of $\rho(H)$ changes sign with increasing temperatures, suggesting a possible crossover of the dominant carrier type from hole-like at low temperatures to electron-like at higher temperatures. By linearly fitting the Hall resistivity curves, the Hall coefficient $R_H$ was extracted and is plotted in Fig. 3d. A pronounced minimum in $R_H(T)$ is observed near 50 K, close to the characteristic temperature identified from the kink in the resistivity at 48 K. This result indicates that the anomaly is accompanied by a change in the carrier response, likely involving a redistribution of carrier contributions or a change in electronic structures.

The magnetic properties were further investigated by the temperature-dependent magnetization measurements. Under $\mu_0 H = 1$ T, the magnetic susceptibility of $Cs_3V_9Te_{13}$ exhibits a monotonic decrease with increasing temperatures under magnetic fields applied along the *c* axis and the *ab* plane (Fig. 3e). A distinct kink appears at $T \approx 48$ K, which is consistent with the resistance measurement. To further analyze the magnetic response of $Cs_3V_9Te_{13}$, the high-temperature susceptibility between 60 and 130 K was fitted with the modified Curie–Weiss law, $\chi = \chi_0 + C / (T - \theta_W)$, where $\chi_0$, $C$, and $\theta_W$ are the temperature-independent susceptibility, the Curie constant and Weiss temperature, respectively. As shown in Fig. S1, $1/(\chi-\chi_0)$ displays an approximately linear temperature dependence within the fitting range. The fit yields a Weiss temperature of $\theta = -61.5$ K for $H // ab$ and $\theta = -47.0$ K for $H // c$. The negative Weiss temperature indicates dominant antiferromagnetic correlations in this compound. The modified Curie–Weiss fit yields an effective moment of 1.73 $\mu_B$ per formula unit for $H // ab$, corresponding to about 0.58 $\mu_B$ per V atom. Such a reduced effective moment may indicate that the V magnetism is at least partly itinerant, rather than arising from fully localized local moments. Furthermore, the ZFC curves measured under different magnetic fields (Fig. 3f) reveal no obvious suppression of the kink, in good agreement with the resistivity data. Combined with the transport anomaly at the same temperature, it reveals a possible phase transition that is robust against the applied magnetic fields.

The resistivity, Hall coefficient, and magnetic susceptibility all exhibit a clear anomaly near 48 K. To further determine whether this anomaly is accompanied by a structural change, the temperature-dependent x-ray diffraction measurements were performed. The diffraction patterns collected at 40 K and 100 K along several crystallographic directions are shown in Fig. S2, including the 0kl, h0l, and hk0 reciprocal-space planes. A comparison of the diffraction patterns collected at 40 K and 100 K reveals no discernible change, suggesting that the anomaly near 48 K is not associated with a structural phase transition. In addition, because a conventional long-range charge-density-wave transition is typically accompanied by lattice modulation and superlattice reflections[27], the absence of clear extra diffraction spots argues against a CDW-like transition with the pronounced structural modulation. Therefore, considering the weak field dependence observed in both resistivity and magnetization, these results suggest that the anomaly at 48 K is more likely associated with a likely electronic and/or magnetic origin.

To further explore the tunability of the electronic state, we carried out the high-pressure transport measurements on $Cs_3V_9Te_{13}$. As shown in Fig. 4, the hump-like character observed at the ambient pressure is rapidly suppressed at 1.5 GPa, and the sample shows a more metallic state, as evidenced by the decrease in resistance upon cooling. Meanwhile, the kink at 48 K seen at the ambient pressure is no longer resolved at 1.5 GPa. With increasing pressures, the metallic character is progressively enhanced up to approximately 16 GPa, but weakens again upon the further compression. Overall, the resistance of $Cs_3V_9Te_{13}$ shows a non-monotonic dependence on the applied pressure, suggesting a complex interplay between the lattice compressions and the electronic states. High-pressure x-ray diffraction measurements are therefore crucial for clarifying whether lattice reconstruction is involved, while the transport measurements at higher pressures are needed to further explore the pressure-dependent evolution of $Cs_3V_9Te_{13}$.

Fig. 5a illustrates the first-principles electronic band structure of $Cs_3V_9Te_{13}$ projected onto the atomic orbitals with the considering spin-orbit coupling (SOC). Our analysis of various magnetic configurations reveals that the ground state is an antiferromagnetic structure characterized by the in-plane ferromagnetism and interlayer antiferromagnetism. Calculations incorporating SOC with different magnetization directions indicate an easy axis along the *a*-direction, yielding a magnetic anisotropy energy of approximately 1.5 meV/f.u. The calculated magnetic moments for the two inequivalent V atoms (V1 and V2) are 0.87 and 1.65 $\mu_B$, respectively. With increasing pressures, these magnetic moments gradually diminish and completely vanish at around 40 GPa. This demonstrates that the pressure can effectively suppresses the intrinsic magnetism of the system, which may be closely related to the pressure-induced changes in the electronical transport properties.

As shown in the projected band structure, the electronic states near the Fermi level ($E_F$) are predominantly contributed by the V atoms. Notably, multiple Dirac points are observed at the K (H) points, while several the van Hove singularities emerge at the M (L) points near $E_F$. In addition, several bands exhibit relatively the flat dispersion throughout the entire Brillouin zone. These features are highly consistent with the tight-binding model results shown in Fig. 1c, further confirming that the Reuleaux-triangle-like lattice hosts intriguing electronic properties analogous to those of the kagome lattice. The Fermi surface depicted in Fig. 5b exhibits cylindrical characteristics, which aligns with the weak band dispersion along the $k_z$ direction observed in Fig. 5a, reflecting the 2D nature of the electronic structure driven by the interlayer van der Waals interactions. Furthermore, the relatively small Fermi surface implies a weak metallicity, which corresponds to the pronounced dip in the electronic density of states (DOS) at the Fermi level. Moreover, the dynamical stability is confirmed by the calculated phonon spectrum in Fig. 5d, suggesting that the experimentally observed phase transition near 48 K is not of a structural nature.

In summary, we have synthesized a new layered vanadium-based compound, $Cs_3V_9Te_{13}$, whose crystal structure features a distinctive Reuleaux-triangle-like vanadium sublattice. The structural analysis shows that the compound has a layered structure with several intertwined atomic sublattices. At ambient pressure, the transport, Hall, and magnetic measurements consistently reveal an anomaly near 48 K, suggesting a possible phase transition. Temperature-dependent x-ray diffraction measurements show no discernible structural change across this temperature range, indicating that the anomaly is not accompanied by any obvious structural transition. Taken together, these results suggest a nonstructural transition of likely electronic and/or magnetic origin. Furthermore, the high-pressure transport measurements reveal a non-monotonic evolution of metallicity, highlighting the strong tunability of the electronic state under the lattice compression. Crucially, our theoretical calculations reveal a quasi-2D kagome-like exotic electronic structure, and identify a dynamically stable antiferromagnetic ground state with the pressure-suppressed magnetism. These results demonstrate $Cs_3V_9Te_{13}$ as a new platform for exploring the interplay between nontrivial lattice geometry, magnetic correlations, and emergent phenomena.

**Single crystal growth of $Cs_3V_9Te_{13}$.** Single crystals of $Cs_3V_9Te_{13}$ were grown from Cs liquid (purity 99.98%), V powder (purity 99.9%) and Te shots (purity 99.999%) via a self-flux method. The mixture was placed into an alumina crucible and sealed in a quartz ampoule under an argon atmosphere. The mixture was heated to 1000 ºC and soaked for 24 h, and subsequently cooled to 600ºC at 2 ºC/h. Finally, the flux was removed by centrifugation at 600 °C, and crystals with regular morphology and metallic luster can be obtained. Due to the high reactivity of the alkali metals, all weighing procedures were carried out in a glovebox under an atmosphere with $O_2$ < 0.01 ppm and $H_2O$ < 0.01 ppm. The stoichiometric ratio of $Cs_3V_9Te_{13}$ was confirmed by the scanning electron microscopy (SEM) and energy-dispersive X-ray spectroscopy (EDS).

**Sample characterization**. XRD patterns were collected using a Rigaku SmartLab SE x-ray diffractometer with Cu Kα radiation (λ = 0.15418 nm) at room temperature. Single-crystal x-ray diffraction and temperature-dependent XRD measurements were carried out using a Bruker D8 diffractometer. SEM and EDS analyses were performed using a HITACHI S5000 with an energy dispersive analysis system Bruker XFlash 6|60. Magnetic susceptibility was determined by a SQUID magnetometer (Quantum Design MPMS XL-1). Electrical resistivity data were collected on a Quantum Design Physical Properties Measurement System (PPMS). The cross-sectional sample along the [120] projection was prepared using a focused ion beam (FIB) system. Atomic-scale STEM imaging was carried out on an aberration-corrected Nion U-HERMES100 dedicated STEM, operated at an acceleration voltage of 60 kV.

**First-principles calculations.** All density functional theory (DFT) simulations were executed via the *VASP* code[28]. We employed the Perdew-Burke-Ernzerhof (PBE) [29] generalized gradient approximation (GGA) to describe the exchange and correlation effects. A plane-wave cutoff energy of 520 eV and a Γ-centered k-point sampling 6×6×8 were chosen to ensure numerical accuracy. The geometry relaxation was performed until the energy and force convergence limits tightly reached $10^{-7}$ eV/atom and 1 meV/Å, respectively. Additionally, phonon properties and vibrational modes were computed using the finite displacement method as implemented in the *phonopy* software [30], utilizing the DFT-derived force constants.

**High-pressure electrical property measurements.** High-pressure electrical transport measurements of $Cs_3V_9Te_{13}$ were performed in the BeCu-type diamond anvil cell (DAC) with 300 μm flat by employing standard four-probe method. The KBr was used as the solid pressure transmitting medium (PTM). The size of single-crystal samples was about 60×20×10 μm$^3$.

**Acknowledgements**

The work is supported by grants from the National Key Research and Development Projects of China (2022YFA1204100), the National Natural Science Foundation of China (62488201, 52572188), the Chinese Academy of Sciences (YSBR-003, YSBR-053). The Innovation Program of Quantum Science and Technology (2021ZD0302700).

**Author Contributions:** H.-J.G. designed the project. H.T.Y. and Z.Z. prepared the samples. Z.Z., R.W.W., and H.T.Y. performed the magnetization measurement and the transport experiments. X.W.Y., and S.G. performed DFT calculations. J.P.S. and J.G.C. performed the pressure related measurement. All authors participated in the data analysis and manuscript writing.

**Competing Interests:** The authors declare that they have no competing interests.

**Data availability**

Data measured or analyzed during this study are available from the corresponding author on reasonable request.

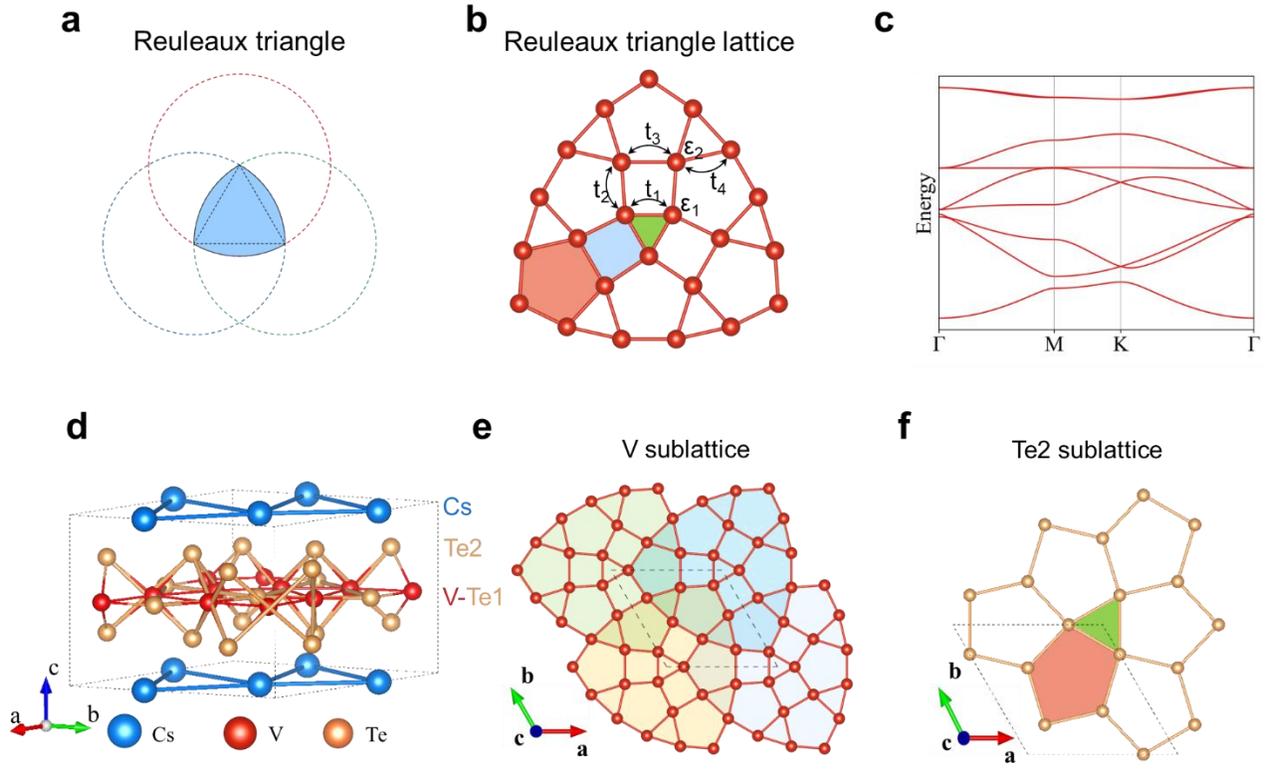

**Fig. 1 Reuleaux-triangle-like structural motif in $Cs_3V_9Te_{13}$.**

(a) Schematic illustration of a Reuleaux triangle. The Reuleaux triangle is shaded in blue, constructed from three equal-radius circular arcs. (b) Reuleaux-triangle-like lattice motif formed by interconnected triangular, quadrilateral, and pentagonal units, highlighted in green, blue, and red, respectively. $t_1$, $t_2$, $t_3$, $t_4$ represent atomic nearest-neighbor hoppings, $\varepsilon_1$ and $\varepsilon_2$ represent the on-site energies of two inequivalent atoms. (c) Tight-binding band structure considering nearest-neighbor hoppings calculated for the Reuleaux-triangle lattice along the high-symmetry path. $t_1$, $t_2$, $t_3$, and $t_4$ are set to -1, -0.9, -0.8, and -0.7 eV, respectively. $\varepsilon_1$ and $\varepsilon_2$ are chosen as 1.5 and 1 eV, respectively. (d) Schematic crystal structure of $Cs_3V_9Te_{13}$, with Cs atoms shown in blue, V atoms in red, and Te atoms in yellow. The dashed lines denote the unit cell. (e) Two-dimensional V sublattice in $Cs_3V_9Te_{13}$, highlighting the Reuleaux-triangle-like network. (f) Two-dimensional Te2 sublattice in $Cs_3V_9Te_{13}$, consisting of an arrangement of edge-sharing pentagons and triangles.

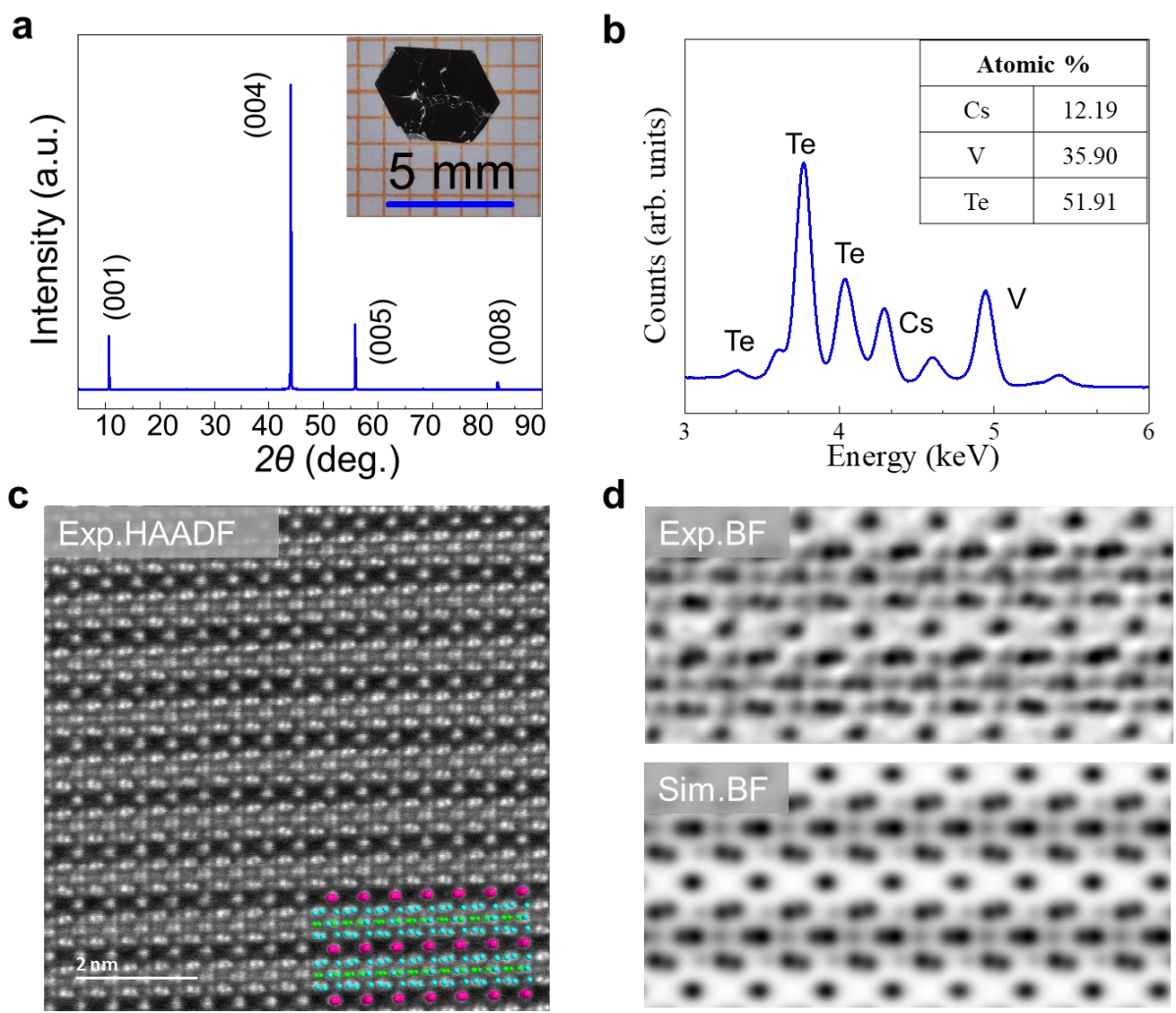

**Fig. 2 Structural characterization and compositional analysis of the $Cs_3V_9Te_{13}$ crystal.**
(a) Single-crystal x-ray diffraction pattern of $Cs_3V_9Te_{13}$, showing only (00l) reflections consistent with its layered structure. The inset shows an optical image of a representative crystal with a regular hexagonal morphology. (b) EDS spectrum and atomic percentage analysis of $Cs_3V_9Te_{13}$, yielding a composition close to the nominal stoichiometry with Cs: V: Te = 3.1: 9: 13.0. (c) Atomic-resolution HAADF-STEM image viewed along the [120] direction, with the corresponding structural model superimposed. (d) Experimental and simulated BF-STEM images along the same projection, showing a good agreement with the structural model.

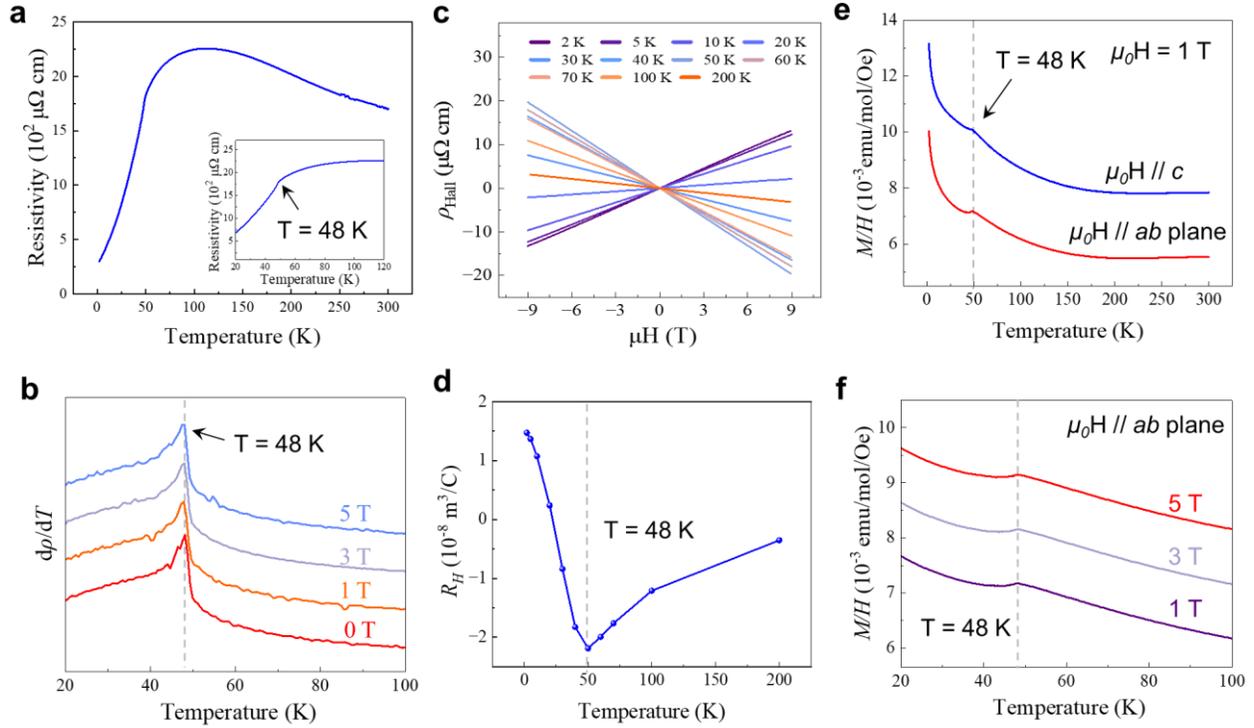

**Fig. 3. Electrical transport and magnetic measurement of $Cs_3V_9Te_{13}$ single crystal.**
**(a)** Temperature-dependent resistivity of $Cs_3V_9Te_{13}$, showing a non-monotonic temperature dependence with a hump-like feature. The inset presents an enlarged view, where a kink at 48 K can be clearly identified. **(b)** Temperature derivative of the resistivity, dρ/dT, under different magnetic fields, highlighting a pronounced anomaly at 48 K with weak field dependence. The curves are vertically offset for clarity. **(c)** Field-dependent Hall resistivity, $ρ_{xy}$, measured at different temperatures. **(d)** Temperature dependence of the Hall coefficient, $R_H$, extracted from linear fits to the Hall resistivity curves. A pronounced minimum is observed, close to the characteristic temperature identified from the resistivity anomaly. **(e)** Magnetic susceptibility, *M/H*, measured under $μ_0H$ = 1 T with the magnetic field applied parallel to the *c* axis and the *ab* plane. Both curves show an anomaly at near 48 K. **(f)** Temperature-dependent *M/H* measured with $μ_0H$ parallel to the *ab* plane under different magnetic fields, indicating that the 48 K anomaly is robust against magnetic fields. The curves are vertically offset for clarity, with a shift of 1 × 10$^{-3}$ emu/mol/Oe between the adjacent curves.

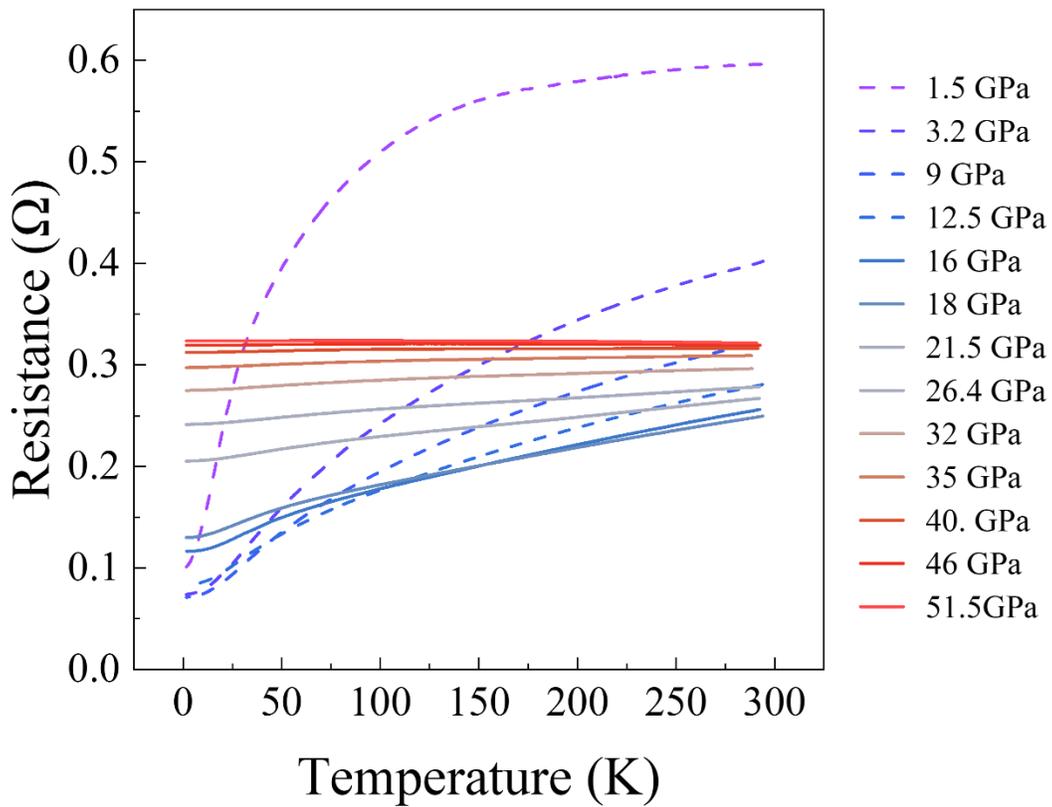

**Fig. 4 Pressure-dependent resistance of Cs$_3$V$_9$Te$_{13}$.**

Temperature-dependent resistance of Cs$_3$V$_9$Te$_{13}$ measured under the different applied pressures. The hump-like anomaly observed at ambient pressure is progressively suppressed under the compression, and the transport behavior shows a non-monotonic evolution with increasing pressures. For clarity, the curves measured from 1.5 to 12.5 GPa are shown as dashed lines, whereas the curves measured at higher pressures are plotted as solid lines.

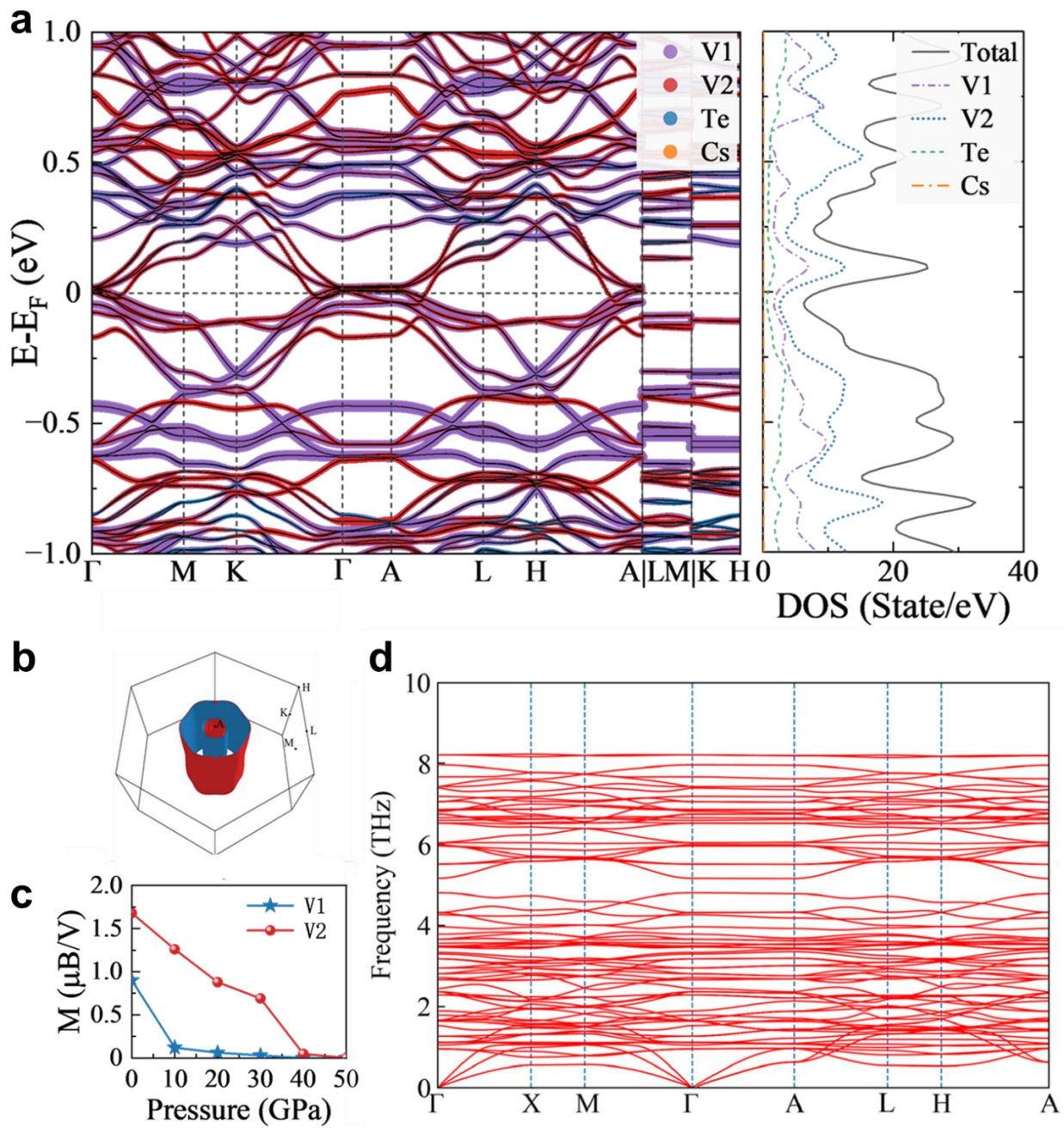

**Figure 5. Calculated electronic properties of $Cs_3V_9Te_{13}$.**
(a) Left panel: Electronic band structure of $Cs_3V_9Te_{13}$ projected onto different atomic orbitals with spin-orbit coupling (SOC). Right panel: Corresponding partial density of states (PDOS) projected onto different atoms. (b) Three-dimensional Fermi surface of $Cs_3V_9Te_{13}$, exhibiting cylindrical characteristics. (c) Evolution of the magnetic moments of V atoms as a function of applied pressures. (d) Calculated phonon spectrum of $Cs_3V_9Te_{13}$ under the ambient pressures.

# Supporting Information

**Table S1. Sample and crystal data for $Cs_3V_9Te_{13}$.**

| Chemical formula | $Cs_3V_9Te_{13}$ |
|---|---|
| Formula weight | 2514 g/mol |
| Temperature | 273(2) K |
| Wavelength | 0.71073 Å |
| Crystal system | hexagonal |
| Space group | P -6 2 m |
| Unit cell dimensions | a = 10.1327(9) Å; b = 10.1327(9) Å; c = 8.2181(9) Å |
| Angles | $\alpha = 90°; \beta = 90°; \gamma = 120°$ |
| Volume | 730.72(15) Å³ |
| Z | 3 |
| Density (calculated) | 5.717 g/cm³ |
| Absorption coefficient | 19.134 mm⁻¹ |
| F(000) | 1048 |

**Table S2. Data collection and structure refinement for $Cs_3V_9Te_{13}$.**

| Item | Value |
|---|---|
| Theta range for data collection | 2.32 to 28.24° |
| Index ranges | -11<=h<=13, -8<=k<=13, -10<=l<=10 |
| Reflections collected | 3933 |
| Independent reflections | 707 [R(int) = 0.0646] |
| Coverage of independent reflections | 100.0% |
| Absorption correction | Multi-Scan |
| Structure solution technique | direct methods |
| Structure solution program | SHELXT 2014/5 (Sheldrick, 2014) |
| Refinement method | Full-matrix least-squares on $(F^2)$ |
| Refinement program | SHELXL-2018/3 (Sheldrick, 2018) |
| Function minimized | $\Sigma\ w(F_o^2 - F_c^2)^2$ |
| Data / restraints / parameters | 707 / 0 / 28 |
| Goodness-of-fit on (F^2) | 1.079 |
| Final R indices | 653 data; (I > 2σ (I)) (R1 = 0.0339), (wR2 = 0.0727) all data: (R1 = 0.0385), (wR2 = 0.0749) |
| Weighting scheme | (w = $1/[\sigma^2(F_o^2) + (0.0313P)^2 + 0.9669P]$) where P=$(F_o^2+2F_c^2)/3$ |
| Absolute structure parameter | 0.42(16) |

**Table S3.** Atomic coordinates and equivalent isotropic atomic displacement parameters (Å$^2$) for Cs$_3$V$_9$Te$_{13}$.

|     | x/a          | y/b          | z/c          | U(eq)      |
|-----|--------------|--------------|--------------|------------|
| Cs  | 0.0          | 0.3632(2)    | 0.0          | 0.0345(5)  |
| V1  | 0.8220(4)    | 0.5334(4)    | 0.5          | 0.0164(7)  |
| V2  | 0.8469(4)    | 0.8469(4)    | 0.5          | 0.0172(10) |
| Te1 | 0.0          | 0.77119(12)  | 0.71975(14)  | 0.0172(3)  |
| Te2 | 0.666667     | 0.333333     | 0.25213(18)  | 0.0195(3)  |
| Te3 | 0.58986(18)  | 0.58986(18)  | 0.5          | 0.0485(8)  |

**Table S4.** Anisotropic atomic displacement parameters (Å$^2$) for Cs$_3$V$_9$Te$_{13}$.

|     | U11        | U22        | U33        | U23        | U13 | U12        |
|-----|------------|------------|------------|------------|-----|------------|
| Cs  | 0.0233(10) | 0.0385(9)  | 0.0365(12) | 0          | 0   | 0.0116(5)  |
| V1  | 0.0125(15) | 0.0120(16) | 0.0242(16) | 0          | 0   | 0.0056(12) |
| V2  | 0.0118(16) | 0.0118(16) | 0.029(3)   | 0          | 0   | 0.0067(19) |
| Te1 | 0.0134(6)  | 0.0152(5)  | 0.0224(6)  | -0.0001(4) | 0   | 0.0067(3)  |
| Te2 | 0.0194(5)  | 0.0194(5)  | 0.0196(7)  | 0          | 0   | 0.0097(3)  |
| Te3 | 0.0079(6)  | 0.0079(6)  | 0.129(3)   | 0          | 0   | 0.0030(7)  |

The anisotropic atomic displacement factor exponent takes the form: $-2\pi^2[h^2 a^{*2} U_{11} + ... + 2hk a^* b^* U_{12}]$

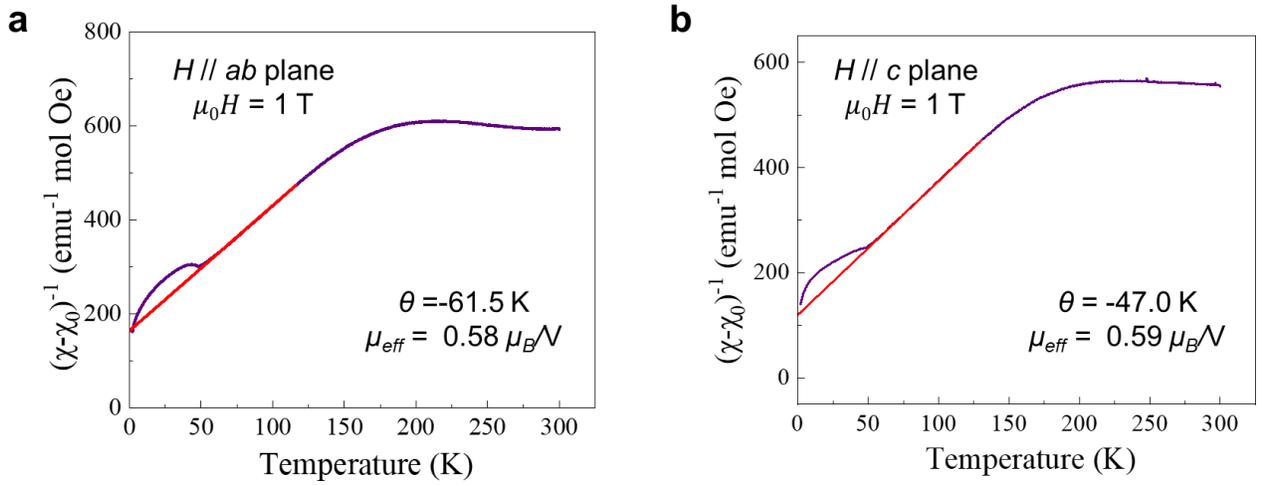

Figure S1. **Curie–Weiss Analysis of the Magnetic Susceptibility.**
(a) Temperature dependence of the inverse magnetic susceptibility measured under $\mu_0H = 1$ T for $H \parallel ab$ plane and $H \parallel c$. (b). The solid red lines are Curie–Weiss fits, yielding $\theta = -61.5$ K and $\mu_{\text{eff}} = 0.58\ \mu_B$ per V atom for $H \parallel ab$ plane, $\theta = -47.5$ K and $\mu_{\text{eff}} = 0.59\ \mu_B$ per V atom for $H \parallel c$.

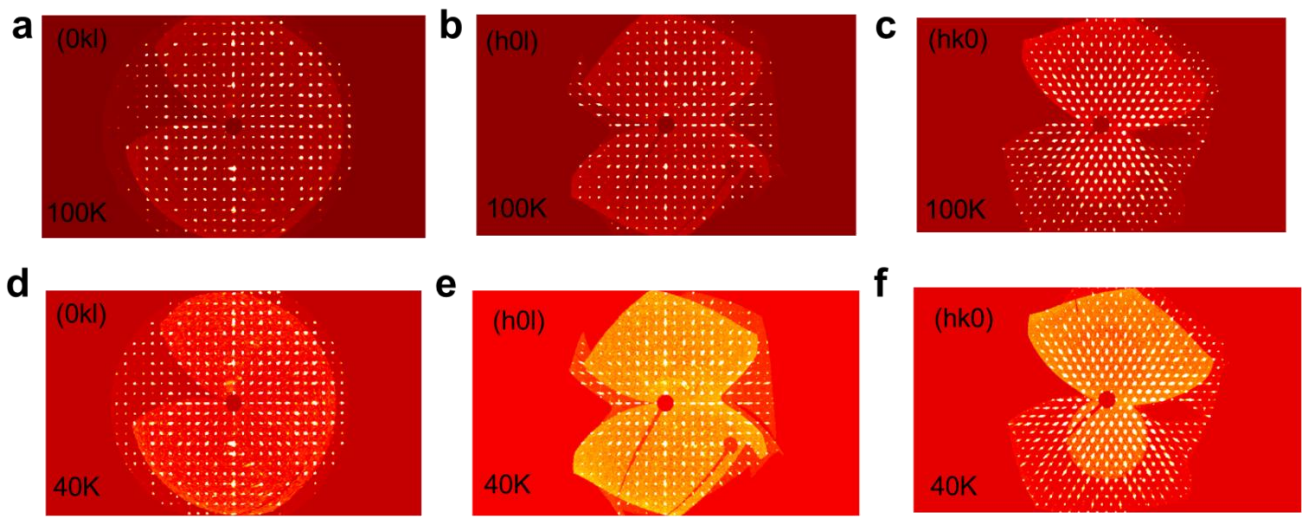

**Figure S2. Temperature-dependent X-ray diffraction patterns of Cs$_3$V$_9$Te$_{13}$ measured along different crystallographic directions.**
(a–c) Diffraction patterns collected at 100 K along three different crystallographic directions. (d–f) Diffraction patterns collected at 40 K along the same directions.